\documentclass[preprint,12pt,authoryear]{elsarticle}

\usepackage{amssymb}
\usepackage{amsmath}

\usepackage{algorithm}
\usepackage{algpseudocode}

\usepackage{multirow}

\usepackage{amsthm}
\usepackage{amssymb}
\usepackage{amsmath}
\usepackage{mathrsfs}
\usepackage{mathtools}

\usepackage{bm}
\usepackage{cite}

\usepackage{graphicx}
\usepackage{subfigure}
\usepackage[usenames]{color}
\usepackage{xcolor}

\usepackage{balance}

\usepackage{enumitem}
\usepackage{makecell}
\usepackage{scalerel}
\usepackage[font={small}]{caption}

\usepackage{accents}

\DeclareMathAlphabet{\mathpzc}{OT1}{pzc}{m}{it}
\DeclarePairedDelimiterX{\norm}[1]{\lVert}{\rVert}{#1}

\newtheorem{lemma}{\bf{Lemma}}

\newtheorem{thm}{Theorem}
\newtheorem{remak}{Remark}

\algnewcommand\Input{\item[\hspace{6pt}\textbf{Input:}]}
\algnewcommand\Output{\item[\hspace{6pt}\textbf{Output:}]}
\algnewcommand\OutputVal{\textbf{output} }
\newcolumntype{L}[1]{>{\raggedright\arraybackslash}p{#1}}
\newcolumntype{C}[1]{>{\centering\arraybackslash}p{#1}}
\newcolumntype{R}[1]{>{\raggedleft\arraybackslash}p{#1}}

\newcommand{\N}{\mathbb{N}}
\newcommand{\R}{\mathbb{R}}
\newcommand{\C}{\mathbb{C}}

\newcommand{\cE}{\mathcal E}
\newcommand{\cP}{\mathcal P}

\newcommand{\bE}{\mathbf E}
\newcommand{\bF}{\mathbf F}
\newcommand{\bK}{\mathbf K}
\newcommand{\bL}{\mathbf L}
\newcommand{\bM}{\mathbf M}

\newcommand{\bR}{\mathbf R}

\DeclareMathOperator{\diag}{diag}
\DeclareMathOperator{\re}{Re}
\DeclareMathOperator{\im}{Im}

\DeclareMathOperator{\trace}{tr}

\journal{EJC}

\begin{document}

\begin{frontmatter}



\title{Distributed Koopman Operator Learning from Sequential Observations\tnoteref{label1234}} 
\tnotetext[label1234]{This research was funded by the Italian Ministry of Enterprises and Made in Italy for the project 4DDS (4D Drone Swarms) under grant no. F/310097/01-04/X56, and the National Natural Science Foundation of China under grant no. 62203053.}



            \author[A1]{Ali Azarbahram} 
\affiliation[A1]{organization={The Department of Electrical Engineering, Chalmers University of Technology},
            city={Gothenburg},
            postcode={412 96}, 
            country={Sweden}}

\author[A2]{Shenyu Liu} 
\affiliation[A2]{organization={The School of Automation, Beijing Institute of Technology},
            country={China}}

\author[A3]{and Gian Paolo Incremona} 
\affiliation[A3]{organization={The Dipartimento di Elettronica, Informazione
e Bioingegneria, Politecnico di Milano},
            city={Milan},
            postcode={20133}, 
            country={Italy}}

\begin{abstract}
This paper presents a distributed Koopman operator learning framework for modeling unknown nonlinear dynamics using sequential observations from multiple agents. Each agent estimates a local Koopman approximation based on lifted data and collaborates over a communication graph to reach exponential consensus on a consistent distributed approximation. The approach supports distributed computation under asynchronous and resource-constrained sensing. Its performance is demonstrated through simulation results, validating convergence and predictive accuracy under sensing-constrained scenarios and limited communication.
\end{abstract}




\end{frontmatter}


\section{Introduction}
\label{Sec. 1}

\textcolor{black}{The challenge of modeling and controlling systems with unknown nonlinear dynamics remains central to many engineering and scientific applications. A foundational idea is to represent nonlinear dynamics through the linear evolution of observables under the Koopman operator, originally introduced by \citep{Koopman1931}. Although the Koopman operator acts on an (in general) infinite-dimensional function space, its spectral objects (eigenvalues, eigenfunctions, and modes) provide a principled way to analyze and approximate nonlinear behavior using linear-algebraic tools. A modern perspective emphasizes how Koopman spectral structure links to global geometry of state space and to representation learning for dynamical systems; see, e.g., the overview in \citep{Mezic2021Notices}.}

\textcolor{black}{Building on this operator-theoretic viewpoint, an extensive body of work has developed practical, data-driven approximations of Koopman representations. Early progress in fluid dynamics connected Koopman spectral analysis to coherent structure extraction and modal decompositions \citep{Rowley2009JFM,Schmid2010JFM}. These ideas were later unified and extended through snapshot-based operator approximation methods, most notably extended dynamic mode decomposition (EDMD), which constructs a finite-dimensional approximation of the Koopman operator from data and a chosen dictionary of observables \citep{Williams2015JNS}. Complementary lines of work have deepened the theoretical foundations and clarified when finite-dimensional Koopman-invariant subspaces exist, as well as how lifting choices affect approximation quality and computational cost \citep{Mezic2005NonlinearDyn}.}

\textcolor{black}{More recently, the Koopman framework has also expanded toward learning spectral information and operator structure in settings where classical eigen-decompositions are challenging, including methods that infer spectral measures and projections directly from measured trajectories \citep{Korda2020ACHA}. Collectively, these developments position Koopman-based modeling as a practical bridge between nonlinear dynamics and scalable linear prediction/control pipelines, while highlighting that performance hinges on the chosen lifting and on how data are collected, distributed, and communicated across sensing agents \citep{Mezic2021Notices,Williams2015JNS}.}

\textcolor{black}{Koopman-based modeling has been increasingly adopted as a practical bridge between nonlinear dynamics and linear prediction/control tools, with demonstrated impact in diverse application domains.} Representative examples include motion-planning formulations built around Koopman linear predictors \citep{gutow2020koopman}, video-driven learning of dynamical evolution for forecasting \citep{comas2021self}, and robust quadrotor control designs that exploit lifted linear structure for improved performance under uncertainty \citep{oh2024koopman}. In dynamic environments, Koopman predictors have also been used to forecast obstacle motion efficiently and to embed these forecasts into collision-avoidance strategies \citep{lu2024vector}.

\textcolor{black}{On the methodological side, a step toward control-oriented Koopman models is the development of linear predictors for nonlinear controlled systems via lifting and EDMD-type regression, yielding models that can be directly used within model predictive control (MPC) while retaining computational structure comparable to linear MPC \citep{korda2018koopmanMPC}. More recently, Koopman learning has been integrated into safety-critical navigation architectures for UAVs operating in dynamic scenes: \citep{azarbahram2025distributed} introduces a Koopman-enhanced distributed switched MPC framework in which a localized Koopman operator is learned online to predict moving-obstacle trajectories and enable scalable, collision-free coordination; complementary work in \citep{bueno2025koopmanUAV} leverages real-time LiDAR measurements to learn Koopman predictors of surrounding moving objects and embeds them in an MPC loop. Together, these results highlight both the versatility of Koopman predictors across perception--prediction--planning pipelines and their suitability for real-time, constraint-aware decision making in dynamic environments.}

The learning of Koopman operators can be formulated as a data fitting problem, typically solved using least squares algorithms~\citep{PCH-VP-GS:13}.
However, as system size and complexity increase, particularly in multi-agent and networked settings, centralized Koopman learning becomes impractical due to communication, privacy, computation, and scalability limitations. Distributed approaches address these challenges by enabling agents to locally estimate Koopman operators using partial observations and limited communication, thereby improving scalability, preserving privacy, and supporting dynamic network conditions.
Implementing such distributed learning requires solving the least squares problem in a distributed manner. Toward this goal, recent advances in distributed algorithms, particularly those in \citep{XW-JZ-SM-MJC:19,YL-CL-BDOA-GS:19,TY-JG-JQ-XY-JW:20,YH-ZM-JS:22}, \textcolor{black}{achieve exponential convergence guarantees.} For broader context, the surveys~\citep{PW-SM-JL-WR:19, YZ-QL:22} offer comprehensive reviews of related methods.

Recent work has explored various approaches for distributed Koopman operator learning. In~\citep{nandanoori2021data}, a block-wise Koopman formulation is proposed for multi-agent systems, but the focus remains on block-structured learning under central data access assumptions. The studied approach in  \citep{mukherjee2022learning} extends this by exploiting graph sparsity and geometric structure. Deep learning-based methods such as~\citep{hao2024distributed222, hao2024distributed} propose distributed deep Koopman algorithms using neural networks for system identification and control, requiring synchronized neural architectures and backpropagation routines across agents. In~\citep{liu2020towards}, a distributed method is developed under state-based decomposability assumptions to ensure convergence. 
\textcolor{black}{More recently,~\citep{azarbahram2025distributedSafe} investigates distributed Koopman learning with an emphasis on perception and safe navigation through information exchange.}

\textcolor{black}{Our work focuses on the distributed solution of a Frobenius-norm-based Koopman approximation problem under temporally fragmented data access across agents.
This setting arises naturally in sensing-constrained scenarios, such as multi-UAV monitoring or satellite-based observation, where agents are unable to continuously record system trajectories due to bandwidth, energy, or sensing limitations (e.g., Landsat~\citep{wulder2012opening}), or multi-UAV sensing and tracking~(e.g., \citep{101007ffgghh, cao2012overview}).
In such environments, agents must observe a shared system sequentially over time, resulting in partial temporal snapshots that are distributed across the network rather than spatially partitioned or centrally available.
While Koopman-based representations are attractive in practice due to their linear structure and compatibility with prediction and control, existing distributed formulations typically rely on block-structured subsystems, synchronized learning architectures, or specialized information exchange mechanisms.}

\textcolor{black}{In contrast, the problem addressed here considers a shared, nonseparable environment observed intermittently by multiple agents, where no single agent has access to a complete trajectory.
The proposed framework enables agents to collaboratively reconstruct a consistent Koopman operator by fusing temporally fragmented observations through local communication over a generic graph.
This formulation is particularly relevant in practical sensing missions, where sequential data acquisition is a constraint rather than a design choice.
By casting the learning problem as a distributed Frobenius-norm minimization with consensus constraints, the proposed algorithm avoids raw data sharing, deep learning architectures, and parametric model assumptions, while admitting a unified matrix-theoretic convergence analysis.}
\textcolor{black}{The main contributions of this paper are summarized as follows:
\begin{itemize}
    \item We formulate a distributed Koopman operator learning problem under temporally fragmented observations. This targets sensing-constrained scenarios where agents sequentially observe a shared nonlinear system without access to complete trajectories or subsystem decompositions. Unlike existing literature that often assumes spatial partitioning or synchronized data access (e.g.,~\citep{hao2024distributed222, azarbahram2025distributedSafe}), our formulation addresses the challenge of reconstructing global dynamics from intermittent, local temporal snapshots.
    \item We establish a theoretical convergence guarantee for the proposed distributed learning framework. We introduce a novel algorithm based on a proportional-integral (PI) consensus law and provide a rigorous matrix-theoretic analysis to prove its stability. Specifically, we demonstrate that the local Koopman estimates across the network reach a consensus and converge exponentially fast to the optimal centralized Frobenius-norm solution, providing explicit design criteria for the algorithm's step-size and gains.
    \item We validate the framework through a practical application to multi-UAV crowd-density monitoring. By simulating a team of UAVs with staggered sensing schedules, we demonstrate that the distributed approach successfully recovers the evolution of a spatially distributed intensity field. The results show that the learned operators achieve high predictive accuracy on unseen data, effectively bridging the gap between local, fragmented sensing and global, high-fidelity forecasting in a resource-constrained multi-agent environment.
\end{itemize}}

\noindent
The remainder of the paper is structured as follows. The introduction continues with notations and graph definitions; Section II presents the problem formulation and preliminaries. Section III details the proposed method. Section IV
illustrates simulation results, and Section V concludes the paper with future directions.

\medskip

\paragraph*{Notations and graph definition}
	We denote by $\N$ the set of non-negative integers, $\R$ the set of real numbers, $\C$ the set of complex numbers, $\R^n$ the $n$-dimensional real space,
	and $\R^{n\times m}$ the space of $n\times m$ real matrices. In particular, $0_{n\times m}$, (resp.~$1_{n\times m}$) denotes the
	$n\times m$-dimensional zero matrix (resp.~all-ones matrix), while $I_n$ represents the $n\times n$ identity matrix. When the dimensions are clear from the context, we remove the subindices. A diagonal matrix composed by elements $a_1, \cdots, a_p$ is denoted by $\diag\left(a_1, \cdots, a_p\right)$, and a block diagonal matrix composed by matrices  $A_1,\cdots, A_p$ is denoted by $\diag\left(A_1,\cdots,A_p\right)$. The Kronecker product of two matrices $A, B$ is denoted by $A \otimes B$, and the transpose of matrix $A$ is denoted by $A^\top$. For square matrix $A\in\R^{n\times n}$, let $\det(A)$ be its determinant, $\trace(A)$ be its trace, and $\Lambda(A)$ be its spectrum (i.e., set of all eigenvalues). For $\lambda\in\C$, let $\re(\lambda),\,\im(\lambda)$ be its real and imaginary parts, respectively.
	For any vector $x\in\R^n$, let $\Vert x\Vert$ be its 2-norm. For any $A\in\R^{m\times n}$, 
	let $\Vert A\Vert_F:=\trace(A^\top A)$ be its Frobenius norm. 

	An undirected graph $G=(\cP,\cE)$ consists of the vertex set $\cP:=\{1,\ldots,p\}$, $\, p\in\N$, and the edge set $\cE\subseteq\cP\times\cP$, such that
	$(i,j)\in \cE$ if and only if $(j,i)\in \cE$. A path is a sequence
	of vertices connected by edges, and the graph $G$ is
	\emph{connected} if there is a path between any pair of
	vertices. For any $i\in \cP$, the set of neighbors of $i$ is
	$N(i):=\{j\in \cP: (i,j)\in \cE\}$. A \emph{Laplacian matrix} $L=[L_{ij}]\in\R^{p\times p}$ of a graph $G$ is given by
	\begin{equation*}
		L_{ij}=\begin{cases}
			-1,&\mbox{if }(i,j)\in \cE, i\neq j,\\
			0,&\mbox{if }(i,j)\not\in \cE, i\neq j,\\
			-\sum_{k\neq i}L_{ik},&\mbox{if } i=j.
		\end{cases}
	\end{equation*}

\section{Preliminaries and Problem Statement}
\label{Sec. 2}

The Koopman operator framework enables the study of nonlinear dynamical systems using linear operators acting on functions. Consider a discrete-time nonlinear system of the form
\begin{align}
x_{k+1} &= f(x_k), \quad x_k \in \mathcal{M} \subseteq \mathbb{R}^q, \label{eq:nonlinear_dynamics}
\end{align}
where \( f: \mathcal{M} \to \mathcal{M} \) is a nonlinear map and \( q \) is the state dimension.
Rather than studying the trajectory of the state \( x_k \) directly, we analyze the evolution of observables \( \psi: \mathcal{M} \to \mathbb{R} \) that belong to a Hilbert space \( \mathcal{H} \) of scalar-valued functions. 
\textcolor{black}{The infinite-dimensional Koopman operator \( \mathcal{K} : \mathcal{H} \to \mathcal{H} \) is defined as
\begin{align}
(\mathcal{K} \psi)(x) = \psi(f(x)), \quad \forall \psi \in \mathcal{H}, \; x \in \mathcal{M}. \label{eq:koopman_def}
\end{align}
Although the system dynamics \( f \) is nonlinear, the Koopman operator \( \mathcal{K} \) is linear in the space of observables.
We select \( n \) scalar observables and assemble them into a vector-valued function:
\begin{align}
\Psi(x) &= [\psi_1(x), \psi_2(x), \dots, \psi_n(x)]^\top \in \mathbb{R}^n, \label{eq:observable_vector}
\end{align}
where \( \Psi: \mathcal{M} \to \mathbb{R}^n \) maps each state \( x \in \mathcal{M} \) to an \( n \)-dimensional feature space \citep{williams2015data}.
}
We then collect \( N \) data samples \( \{x_k, f(x_k)\}_{k=1}^N \) and define the data matrices
\begin{align}
X &= [\Psi(x_1), \Psi(x_2), \dots, \Psi(x_N)] \in \mathbb{R}^{n \times N}, \label{eq:data_X} \\
Y &= [\Psi(f(x_1)), \Psi(f(x_2)), \dots, \Psi(f(x_N))] \in \mathbb{R}^{n \times N}. \label{eq:data_Y}
\end{align}

To facilitate computation, the EDMD method~\citep{williams2015data} projects the action of \( \mathcal{K} \) onto the span of the selected observables. This results in a matrix \( K \in \mathbb{R}^{n \times n} \) that approximates the Koopman operator over the chosen feature space and governs
the evolution of lifted states in the observable space.
\textcolor{black}{Hence, the finite-dimensional approximation of the Koopman operator \( \mathcal{K} \) via EDMD seeks a matrix \( K \in \mathbb{R}^{n \times n} \) that best satisfies}
\begin{align}
Y &\approx K X. \label{eq:koopman_matrix_approx}
\end{align}
This leads to the following least-squares optimization using the Frobenius norm:
\begin{align}
K^* &= \arg\min_{K \in \mathbb{R}^{n \times n}} \left\| Y - K X \right\|_F^2. \label{eq:frobenius_minimization}
\end{align}

\noindent
In a distributed setting, each agent \( i \in \{1, \ldots, p\} \) collects its own segment of the system observations, resulting in local data matrices \( Y_i \in \mathbb{R}^{n \times m_i} \) and \( X_i \in \mathbb{R}^{n \times m_i} \), where \( m_i \) denotes the number of temporal snapshots available to agent \( i \). These data blocks represent lifted observables evaluated at the agent's local state transitions. Such partitioning naturally arises in sequential observation frameworks, where agents take turns sensing the system over time, or when access to data is constrained by communication, privacy, or coverage limitations. To construct a distributed Koopman approximation, the locally collected data are aggregated to form the global matrices
\[
Y = \begin{bmatrix} Y_1 & Y_2 & \cdots & Y_p \end{bmatrix}, \quad 
X = \begin{bmatrix} X_1 & X_2 & \cdots & X_p \end{bmatrix},
\]
with \( Y,\, X \in \mathbb{R}^{n \times N} \) and \( \sum_{i=1}^{p} m_i = N \). This structure enables decentralized computation while preserving the full representation of the system's evolution.

    From the problem formulation, the $i$-th agent only knows $Y_i, X_i$.
    Because $\Vert Y-KX\Vert_F^2~=\sum_{i=1}^p\Vert Y_i-KX_i\Vert_F^2$, we could study the distributed problem that each agent aims to find a solution $K_i\in\R^{n\times n}$ for the local problem $\min_{K_i}\Vert Y_i-K_iX_i\Vert_F$, subject to the constraint that all $K_i$ must be equal. Furthermore, denote
		\begin{align}
			\bK:&=\begin{bmatrix}
				K_1&K_2&\cdots&K_p
			\end{bmatrix}\in\R^{n\times np},\label{def:bK}\\
			\bm X:&=\diag(X_1,X_2,\cdots,X_p)\in\R^{np\times N},\label{def:bB}\\
			\bL:&=L\otimes I_n\in\R^{np\times np},\label{def:bL}
		\end{align}
		where $L$ is the Laplacian matrix of the communication graph $G$. 
        We further have that \( \Vert Y - KX \Vert_F^2 = \sum_{i=1}^p \Vert Y_i - K X_i \Vert_F^2 = \Vert Y - \bK \bm X \Vert_F^2 \), subject to the constraint \(K=K_1 = K_2 = \cdots = K_p \), {\color{black}that is, $\bK=1_{1\times p}\otimes K$, which can be equivalently expressed as \( \bK \bL = 0 \) since the communication graph \( G \) is connected.}
        Hence, to solve \eqref{eq:frobenius_minimization} by a distributed method in this scenario, we can equivalently study the optimization problem
		\begin{subequations}
		\begin{align}
			\min_{\bK\in\R^{n\times np}}&\frac{1}{2}\Vert Y-\bK \bm X\Vert_F^2, \label{op_1}\\
			&\mbox{subject to } \bK\bL=0. \label{op_2}
		\end{align}	
		\end{subequations}

\section{Main Results}
 \label{Sec. 3} 

	We propose to solve the problem \eqref{eq:frobenius_minimization} by a discrete-time distributed algorithm. Essentially, the agents alternate between communication and computation, such that they run the following updating law
		\begin{subequations}\label{algo1}
			\begin{align}
				K_i^+ &= K_i - \alpha    \Big((K_iX_i-Y_i)X_i^\top  \nonumber \\
                &~~~~~~~~~~~~~~~~~ +k_P\sum_{j\in N(i)} ( K_i-K_j) + k_I R_i    \Big),				\label{algo1_1}\\		R_i^+&=R_i+\alpha\sum_{j\in N(i)} ( K_i-K_j). \label{algo1_2}
			\end{align}
		\end{subequations}
		Here, $\alpha>0$ is a fixed step-size, $k_p,\, k_I>0$ are some tunable gains, $K_i\in\R^{n\times n}$ is the $i$-th agent's individual guess of the optimum of \eqref{eq:frobenius_minimization}, $R_i\in\R^{n\times n}$ is another internal augmented state variable owned by the $i$-th agent for facilitating the convergence. The distributed algorithm for Koopman operator learning is summarized in Algorithm~\ref{algo}.
     The update law is similar to the PI-control studied in \citep{SY-JW-QL:19}, and we provide some intuitive interpretations which help understand Algorithm~\ref{algo}.
    \begin{itemize}
   \item \textcolor{black}{The term \( (Y_i - K_i X_i) X_i^\top \) corresponds to the negative gradient of the local cost \( \frac{1}{2}\|Y_i - K_i X_i\|_F^2 \), indicating that each agent minimizes its own Koopman approximation error; equivalently, the update law is expressed as a gradient-descent step using the gradient \( (K_i X_i - Y_i) X_i^\top \) in \eqref{algo1_1} .}
    \item The term \( \sum_{j \in N(i)} (K_i - K_j) \) in \eqref{algo1_2} acts as a proportional diffusion term, promoting consensus among neighboring agents.
    \item Given \( R_i(0) = 0 \), the update in \eqref{algo1_2} shows that \( R_i \) accumulates the diffusion term over time, functioning as an integral feedback that enforces consensus across agents.
\end{itemize}

\begin{algorithm}[t]
	\caption{Distributed Koopman Operator Learning}\label{algo}
	\begin{algorithmic}[1]
		\Require $k_P, k_I$ and $\alpha$ satisfying $\alpha<\alpha_{\max}$ from \eqref{def:alpha_max}.
		\State Initialize arbitrary $K_i(0)\in\R^{n\times n}$, $R_i(0)=0_{n\times n}$ for all $i\in\cP$.
		\Loop{ for $t=0,1,\ldots,t_{\max}-1$, the $i$-th agent, $i\in\cP$}
		\State Broadcast $K_i(t)$ to its neighbors.
		\State Compute $K_i(t+1), R_i(t+1)$ according to \eqref{algo1}.
		\EndLoop
		\Ensure $K_i(t_{\max})$, $i\in\cP$.
	\end{algorithmic}
\end{algorithm}

    The convergence of Algorithm~\ref{algo} is related to the eigenvalues of the following matrix
	\begin{equation}\label{def:bM}
		\bM:=\begin{bmatrix}
			-\bm X \bm X^\top-k_P\bL& \bL\\
			-k_II_{np}& 0
		\end{bmatrix}\in \R^{2np\times 2np},
	\end{equation}
	where  $\bm X,\bL$ are defined in \eqref{def:bB} and \eqref{def:bL}. Note that $\bM$ depends on the gains $k_P,\,k_I$. The following theorem provides the design criteria of these parameters and the step-size $\alpha$ which guarantees the convergence of Algorithm~\ref{algo}, and relates the convergence rate to these parameters.

	\begin{thm}\label{thm:col} 
		Suppose the communication graph $G$ is undirect and connected. For any gains $k_P,k_I>0$, there exists
		\begin{equation}\label{def:alpha_max}
			\alpha_{\max}:=-\max_{\lambda\in\Lambda(\bM)\backslash\{0\}}\frac{2\re(\lambda)}{|\lambda|^2}>0,
		\end{equation}
		such that as long as the step size $\alpha<\alpha_{\max}$, the $K_i$ components of the algorithm will reach consensus and converge to an optimal solution of the problem~\eqref{eq:frobenius_minimization}. Furthermore, the convergence is exponential with rate $\rho>\rho_{\max}$, where
		\begin{equation}\label{def:rho_max}
			\rho_{\max}:=\max_{\lambda\in\Lambda(\bM)\backslash\{0\}}\sqrt{1+2\alpha\re(\lambda)+\alpha^2|\lambda|^2}.
		\end{equation}
	\end{thm} 

The proof of Theorem~\ref{thm:col} is provided in the Appendix. We note that the upper bound $\alpha_{\max}$ in \eqref{def:alpha_max} on the step size must be computed using the matrix $\bM$, which depends on the centralized quantities $X, L$. Consequently, $\alpha_{\max}$ cannot be determined a priori in a distributed setting. Our convergence guarantee for Algorithm~\ref{algo} should be interpreted qualitatively: provided the step size is sufficiently small, the algorithm converges exponentially fast to an optimal solution of \eqref{eq:frobenius_minimization}. Future work will explore the development of distributed algorithms that either incorporate adaptive step sizes or eliminate the need for step-size tuning altogether.

\section{Simulation results}
\label{Sec. 4} 

\textcolor{black}{
We consider a team of UAVs tasked with monitoring the evolution of a spatially distributed crowd-density field over time. 
Although each UAV is equipped with a camera capable of observing the entire region of interest, continuous high-rate sensing is impractical due to energy limitations, camera usage constraints, and onboard processing costs. 
As a result, the UAVs operate under a sequential sensing protocol, where data acquisition is staggered across agents over a finite time horizon. Specifically, the sensing task is distributed over $k \in \{1,\dots,N\}$ discrete time steps such that each UAV records only a subset of the overall trajectory, while the collective measurements cover the full sequence of snapshots.
This setting naturally leads to temporally fragmented data across agents, without any single UAV having access to the complete dataset.
}

\begin{figure}[!b]
	    \centering
	    \includegraphics[trim=0.0cm 0.0cm 0.0cm 0.0cm, clip, width=0.9\textwidth]{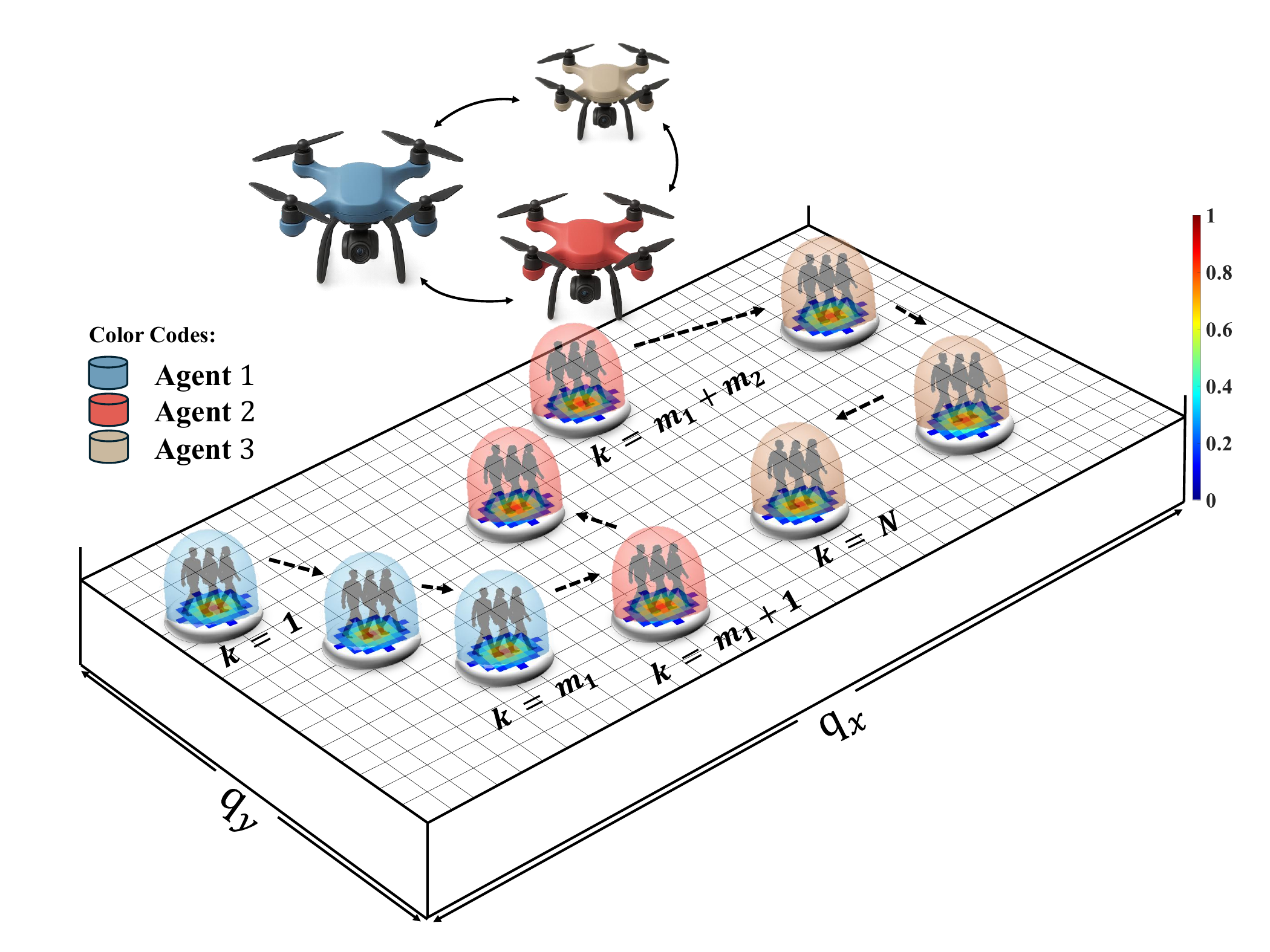}
	    \captionsetup{font=normalsize}
        \caption{\textcolor{black}{Conceptual illustration of sequential sensing and distributed Koopman learning.}}
	    \label{FigureGraphical_Sequen}
	\end{figure}

\textcolor{black}{We consider $p=3$ UAVs connected through a circular, undirected communication graph.
The UAVs acquire measurements in sequence: UAV~1 collects the first $m_1$ snapshots, UAV~2 collects the subsequent $m_2$ snapshots, and UAV~3 collects the remaining $m_3$ snapshots, forming one full sensing cycle with $N = m_1 + m_2 + m_3$ total samples.
Each agent processes only its locally acquired data and participates in the distributed update law~\eqref{algo1} through neighbor-to-neighbor communication.
}

\textcolor{black}{A conceptual illustration of this sequential sensing process is shown in Figure~\ref{FigureGraphical_Sequen}.
The monitored environment is modeled as a two-dimensional grid of size $q_x \times q_y$, yielding a global state dimension $q = q_x q_y$ in~\eqref{eq:nonlinear_dynamics}. In the simulations, we set $q_x = q_y = 20$, resulting in $q = 400$ spatial states.
At each sensing instant, the active UAV observes the full $20 \times 20$ intensity field, but only for its assigned time window in the sensing schedule.
Each UAV acquires $m_i = 3$ snapshot pairs before handing off sensing responsibility to the next agent, resulting in $N = 9$ total samples over one cycle. 
In addition to the snapshots used for learning, we reserve a subsequent sequence of nine future snapshots that are not used during training and are instead employed to evaluate multi-step prediction performance of the Koopman operator learned from the sequentially acquired data.
The evolving crowd generates normalized intensity fields taking values in $[0,1]$, which are visualized using color maps to represent spatial density variations. Shaded regions also indicate which UAV is active at each time interval. For example, a blue-highlighted 
dome from \(k=1\) to \(k=m_1\) denotes measurements acquired by UAV~1.}

\textcolor{black}{The intensity maps constitute the system state $x_k$ and serve as inputs to the Koopman learning procedure.
For lifting, all agents employ a shared observable structure that directly vectorizes the spatial intensity field, yielding a lifted state dimension $n = 400$.
The resulting lifted data are used to construct the local matrices $(X_i, Y_i)$ for each agent, which are then incorporated into the distributed Koopman learning algorithm. The results presented below demonstrate that, despite the absence of centralized data aggregation and the presence of sequentially partitioned observations, the proposed distributed method enables all agents to collectively recover a Koopman operator that closely matches the centralized solution.}

\begin{figure}[!t]
    \centering
    \includegraphics[trim=0.0cm 0.0cm 0.0cm 0.0cm, clip, width=0.9\textwidth]{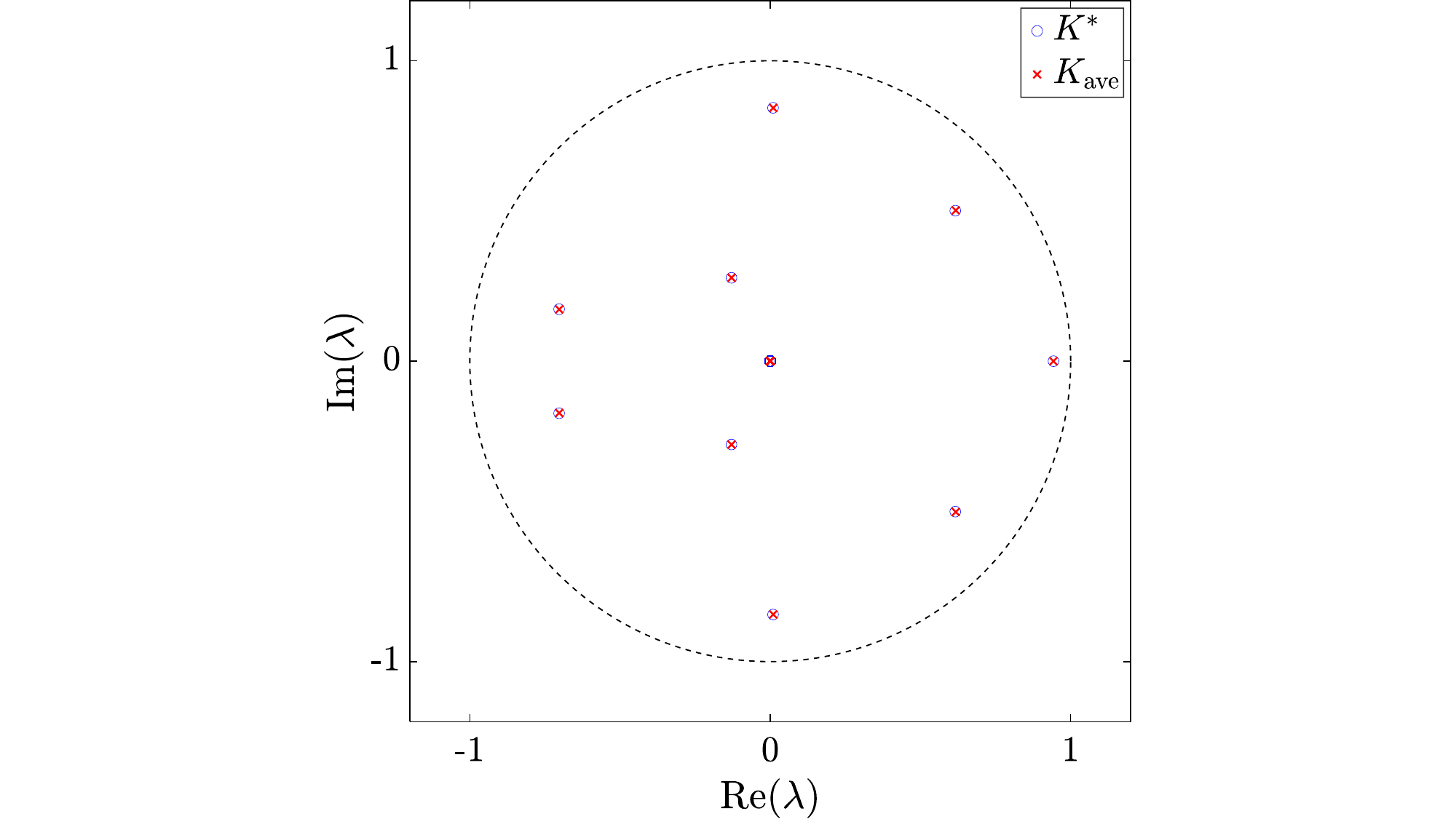}
    \captionsetup{font=footnotesize}
    \caption{
        {\textcolor{black}{Spectral comparison between $K_{\mathrm{ave}}$ and $K^*$.}}
    }
   \label{fig:eigenvalue_comparison}
\end{figure}

\begin{figure}[!t]
    \centering
\includegraphics[trim=0.0cm 0.0cm 0.0cm 0.0cm, clip, width=0.9\textwidth]{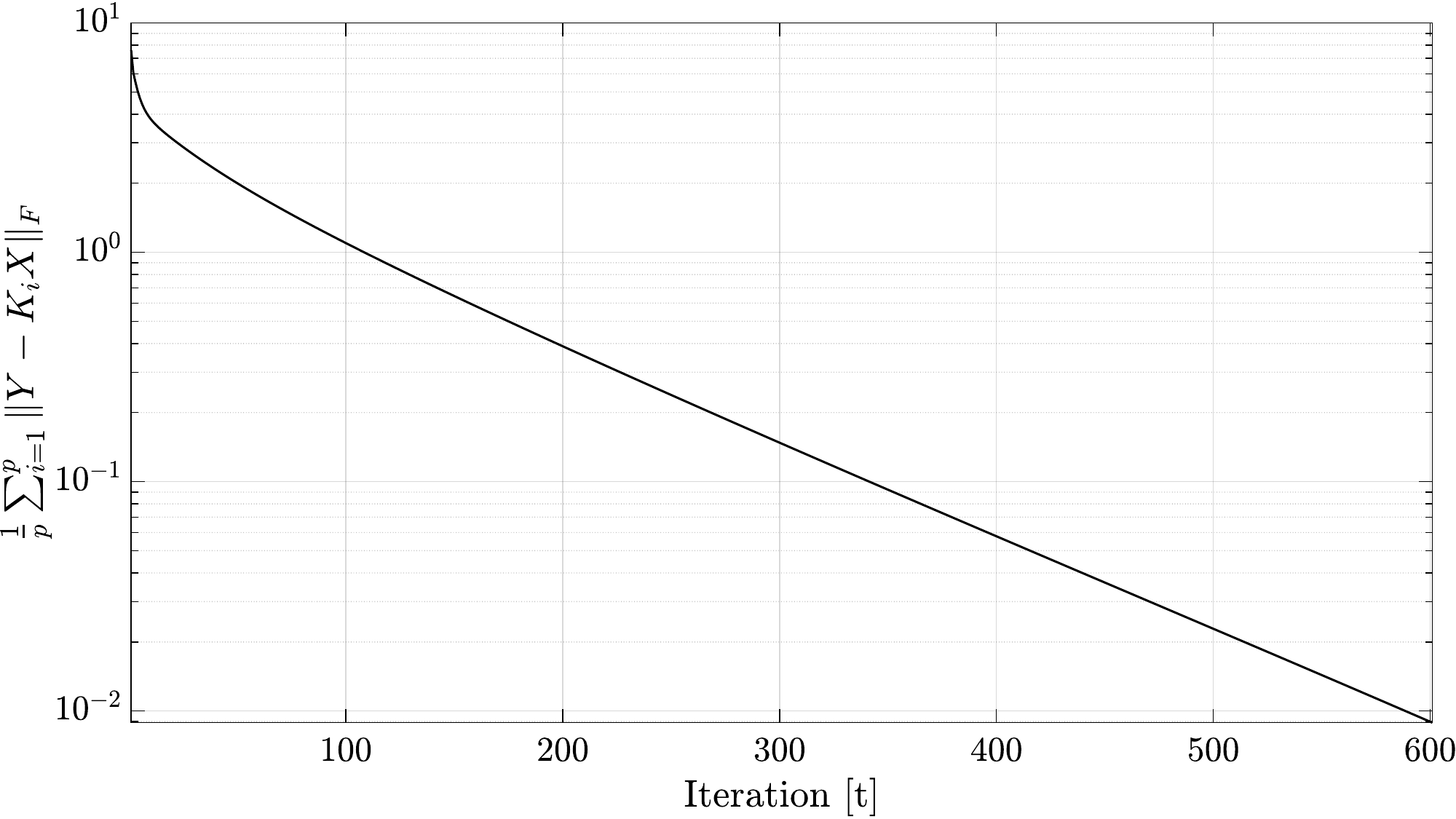}
    \captionsetup{font=footnotesize}
    \caption{
        {\textcolor{black}{Prediction error of distributed Koopman operators.}}
    }
    \label{fig:koopman_fit}
\end{figure}

Each agent runs $600$ iterations of the distributed update, using $k_P = 150$, $k_I = 50$, and $\alpha = 0.5\alpha_{\max}$ with $\alpha_{\max} = 0.03$ from~\eqref{def:alpha_max}. This ensures exponential convergence with $\rho_{\max} = 0.96$ as per~\eqref{def:rho_max}.
 Let $K^*$ be the centralized solution to~\eqref{eq:frobenius_minimization} using $(X,Y)$, and $K_i$ the local estimate from agent $i$. Define the average distributed operator as $K_{\mathrm{ave}} := 1/p \sum_{i=1}^p K_i$.
Figure~\ref{fig:eigenvalue_comparison} shows that $K^*$ and $K_{\mathrm{ave}}$ share dominant eigenvalues near the unit circle, indicating accurate capture of persistent dynamics. Eigenvalues near the origin in $K_{\mathrm{ave}}$ reflect suppressed modes from limited observability and averaging, contributing to robustness.
\textcolor{black}{We should note that the constraint~\eqref{op_2} enforces agreement only at optimality, i.e., at convergence one has
$K_1 = K_2 = \cdots = K_p$.
However, during the transient evolution of Algorithm~\ref{algo}, the local iterates $K_i(t)$ are generally not identical.
For this reason, we define
$K_{\mathrm{ave}}(t)$
as a compact network-level representative of the distributed estimate, which is convenient for visualization and for comparing the collective behavior of the distributed algorithm against the centralized solution $K^*$.
As $t\to\infty$, the consensus error vanishes and $K_{\mathrm{ave}}(t)$ coincides with each $K_i(t)$.}

\begin{figure}[!t]
    \centering
\includegraphics[trim=0.0cm 0.0cm 0.0cm 0.0cm, clip, width=0.9\textwidth]{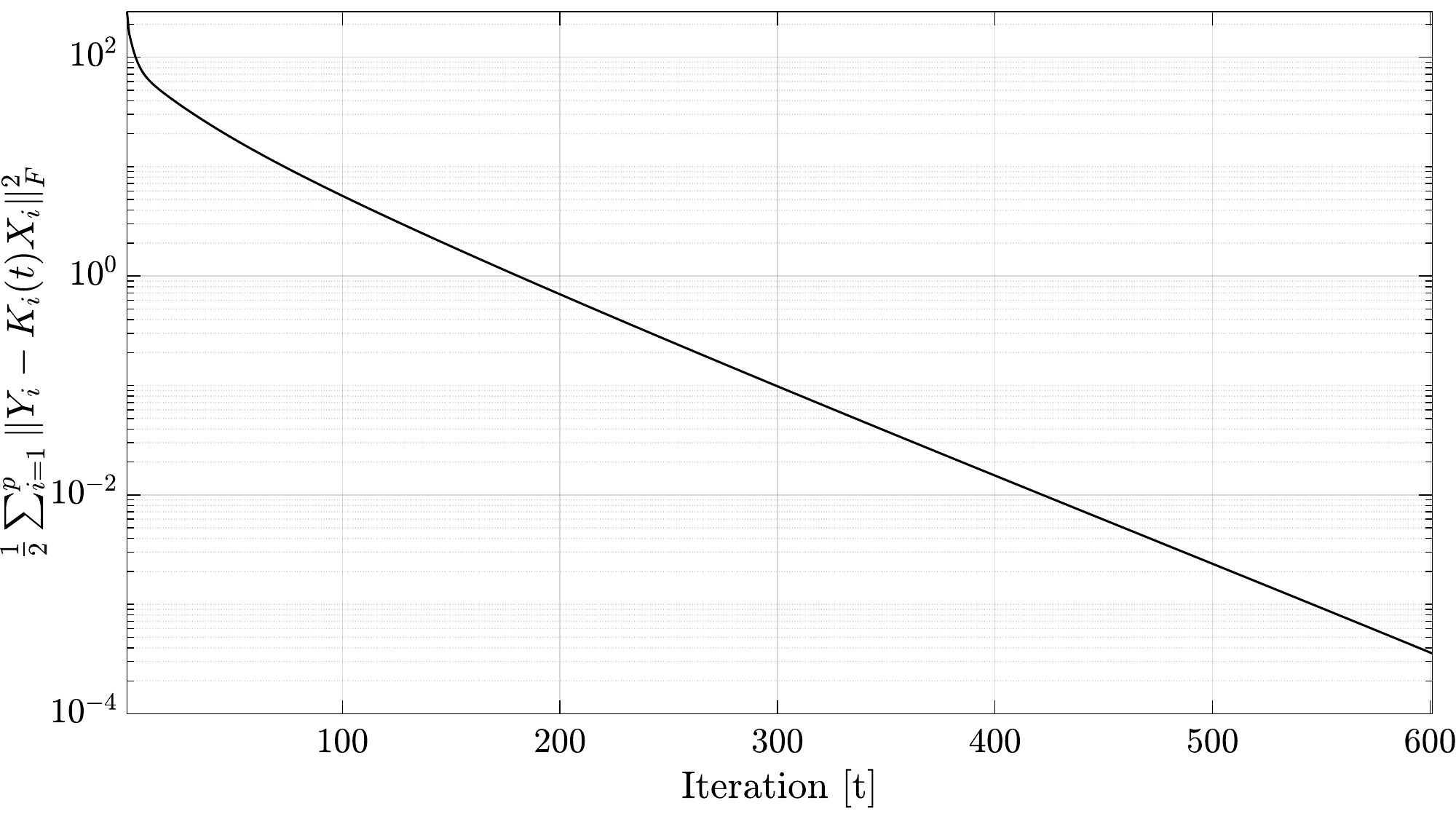}
    \captionsetup{font=footnotesize}
    \caption{
        {\textcolor{black}{Convergence of the distributed Koopman learning objective}}
    }
    \label{fig:koopman_objective}
\end{figure}

\begin{figure}[!t]
    \centering
\includegraphics[trim=0.0cm 0.0cm 0.0cm 0.0cm, clip, width=0.9\textwidth]{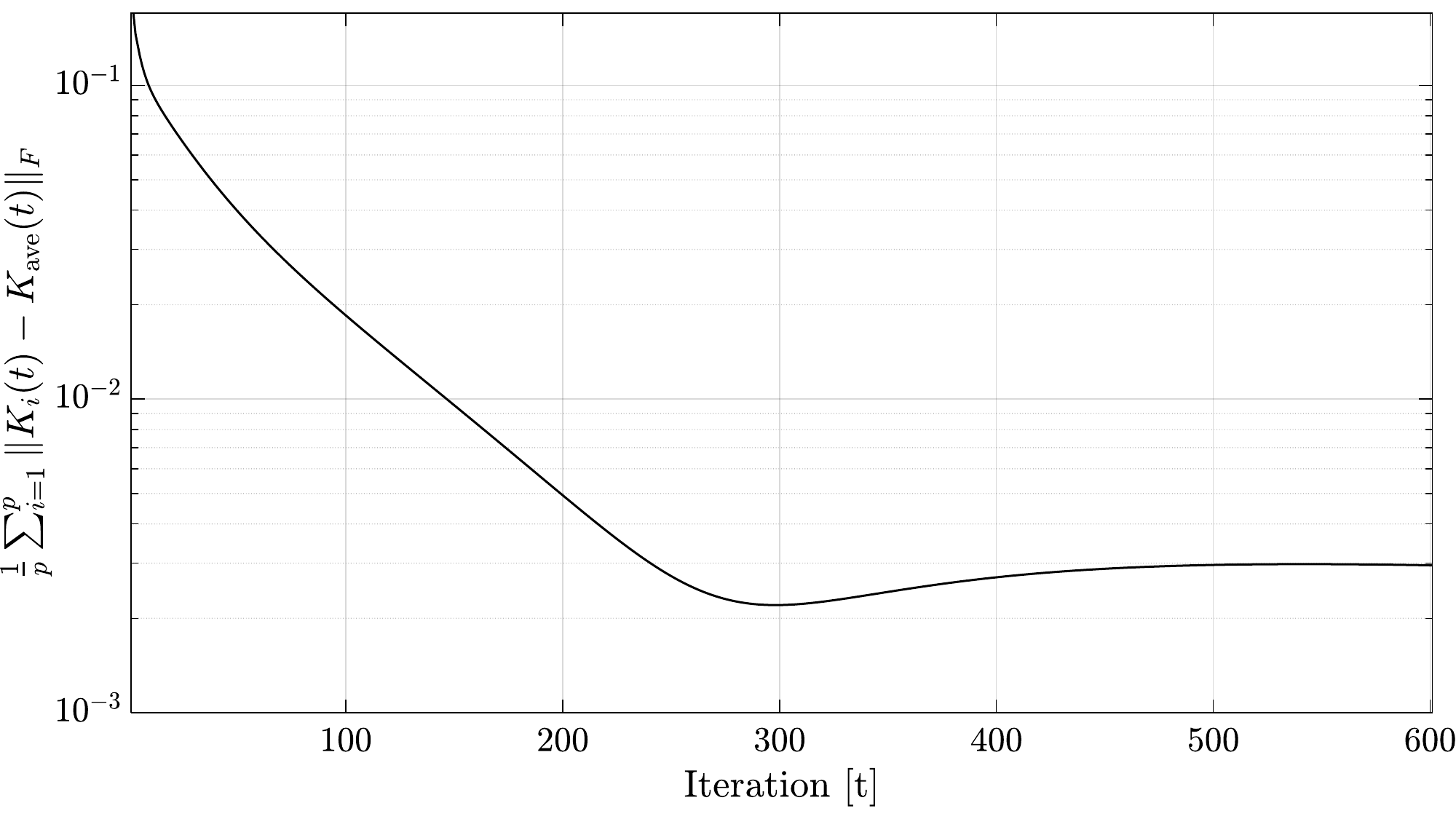}
    \captionsetup{font=footnotesize}
    \caption{
        {\textcolor{black}{Consensus among distributed Koopman operators.}}
    }
    \label{fig:koopman_consensus}
\end{figure}


%
To assess accuracy, we compute $1 / p \sum_{i=1}^p \left\| Y - K_i(t) X \right\|_F$, measuring how well each distributed Koopman operator $K_i$ generalizes to the full dataset $(X,Y)$. Despite being trained on local data, agents achieve low reconstruction error over time, as shown in Figure~\ref{fig:koopman_fit}, indicating alignment with global measurements and inter-agent consensus.
\textcolor{black}{Figure~\ref{fig:koopman_objective} illustrates the evolution of the distributed objective function
$\frac{1}{2}\sum_{i=1}^p \|Y_i - K_i(t)X_i\|_F^2$ over the algorithm iterations.
The results highlight the rapid decrease of the aggregate fitting error, confirming
the exponential convergence behavior predicted by Theorem~\ref{thm:col}.}

\begin{figure}[!t]
    \centering
\includegraphics[trim=0.0cm 0.0cm 0.0cm 0.0cm, clip, width=0.9\textwidth]{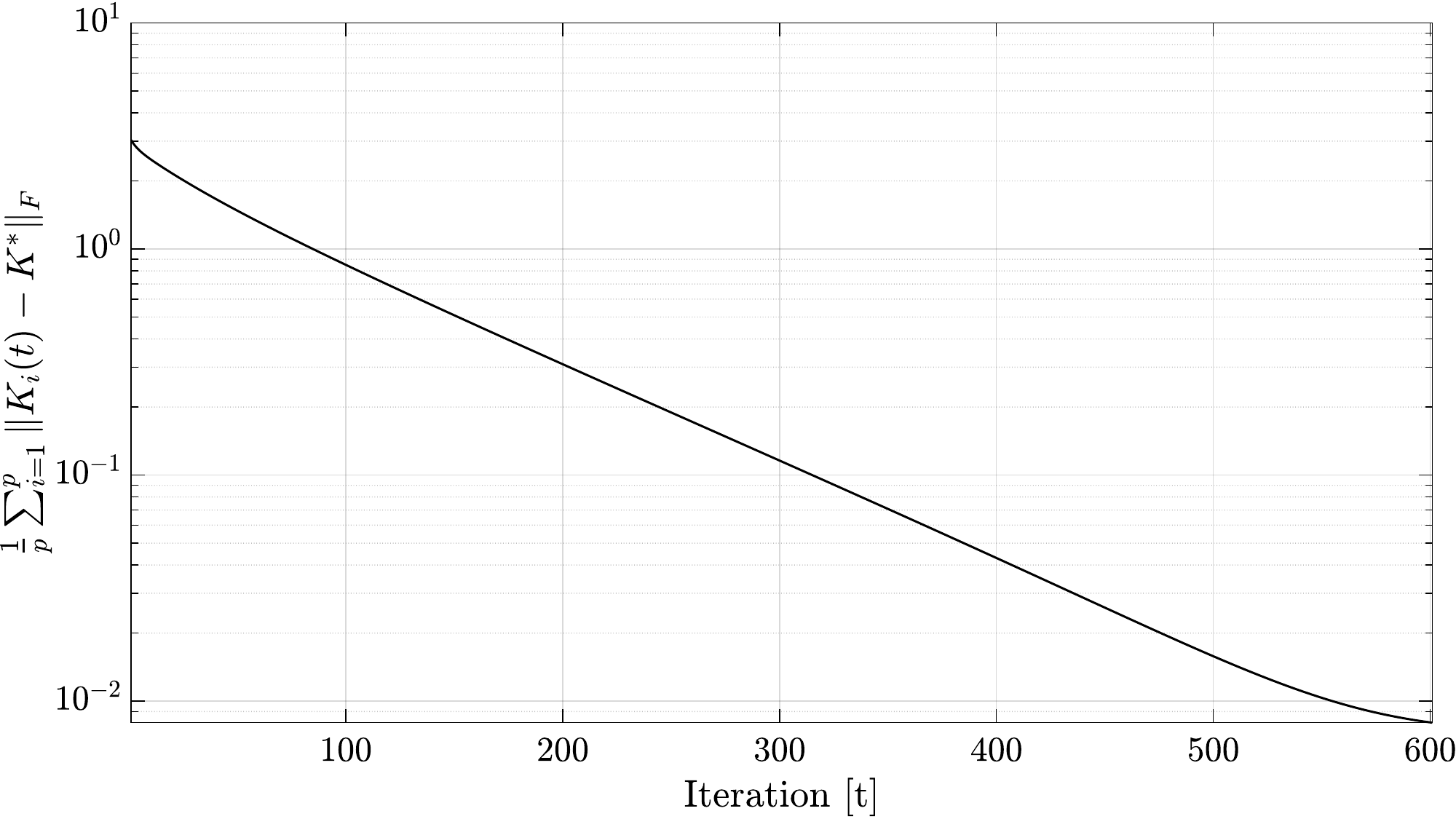}
    \captionsetup{font=footnotesize}
    \caption{
        {\textcolor{black}{Convergence to the centralized Koopman solution.}}
    }
    \label{fig:koopman_central_distance}
\end{figure}

\textcolor{black}{Figure~\ref{fig:koopman_consensus} reports the evolution of the consensus error among agents, quantified by
$\frac{1}{p}\sum_{i=1}^p \|K_i(t)-K_{\mathrm{ave}}(t)\|_F$.
The monotonic decay indicates that the locally learned Koopman operators progressively align through neighbor communication.
This behavior confirms that the distributed update law effectively enforces agreement across agents despite temporally fragmented observations. Figure~\ref{fig:koopman_central_distance} also illustrates the distance between the distributed Koopman operators and the centralized solution $K^*$.
The quantity $\frac{1}{p}\sum_{i=1}^p \|K_i(t)-K^*\|_F$ decreases steadily over the iterations, demonstrating that the distributed learning process converges not only to consensus, but also toward the optimal centralized least-squares solution.}

\begin{figure}[!t]
    \centering
\includegraphics[trim=0.0cm 0.0cm 0.0cm 0.0cm, clip, width=0.9\textwidth]{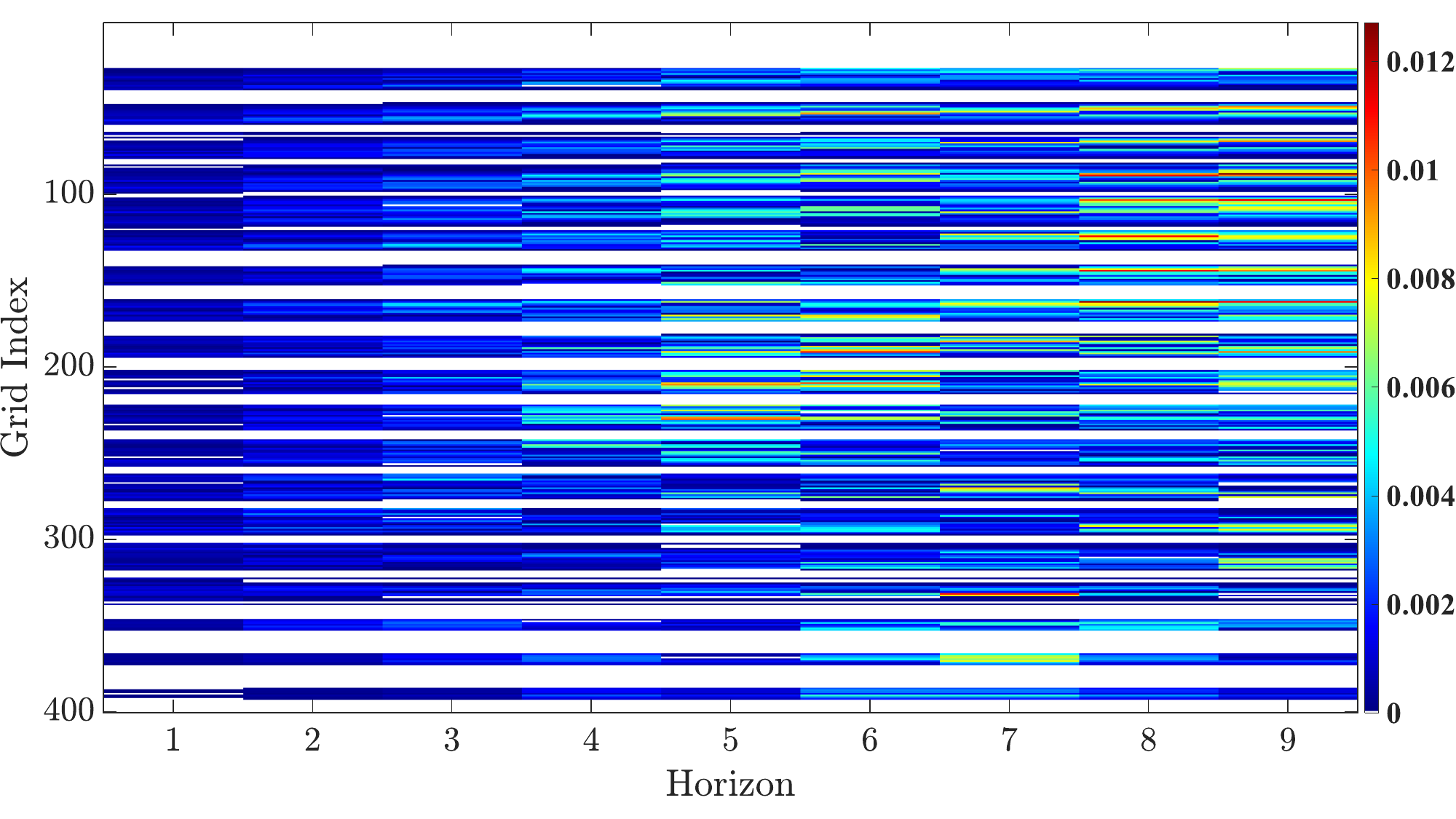}
    \captionsetup{font=footnotesize}
    \caption{
        {\textcolor{black}{Heatmap of distributed Koopman prediction error.}}
    }
    \label{fig:koopman_error_heatmap}
\end{figure}

\textcolor{black}{To assess predictive performance beyond the training horizon, we evaluate the learned Koopman operator on a reserved sequence of snapshots that are not used during the learning phase.
Specifically, after estimating the distributed Koopman operator from the first $N=9$ sequentially acquired samples, an additional set of nine future snapshots is held out and used exclusively for multi-step prediction.
Figure~\ref{fig:koopman_error_heatmap} reports the absolute prediction error between the true system evolution and the distributed Koopman-based forecasts over this future horizon.
The error is shown across both time and spatial coordinates, providing a detailed view of how prediction accuracy evolves as the horizon increases.
The observed errors remain consistently small (on the order of $10^{-2}$) and exhibit coherent spatial structure, indicating stable propagation of prediction accuracy rather than error accumulation.
}

\textcolor{black}{To provide a direct and interpretable comparison between centralized and distributed predictions, we report in Table~\ref{tab:prediction_error_ablation} a time-series evaluation of the prediction error over the reserved (unseen) snapshots.
The centralized Koopman model and the distributed models are used to predict the system evolution over a future horizon, and at each step the error is computed as the $\ell_2$-norm of the difference between the predicted intensity field and the ground-truth snapshot.
The table reports results for both the proposed distributed algorithm with proportional--integral (PI) consensus and a proportional-only (P-only) variant obtained by removing the integral state, which is included as an ablation study to assess the role of the integral term.
The results show that both distributed variants yield stable multi-step predictions on unseen data, indicating that the learned Koopman operators generalize beyond the training snapshots.
While the P-only variant exhibits comparable prediction accuracy over the tested horizons, the PI-based scheme consistently attains lower errors and provides improved agreement with the centralized least-squares solution, in line with the theoretical analysis.
Together, these results highlight the practical forecasting capability of the proposed distributed Koopman learning framework under decentralized and sensing-constrained data acquisition.}

\begin{table}[!t]
\centering
\caption{\textcolor{black}{Time-series prediction error on reserved snapshots.}}
\label{tab:prediction_error_ablation}
\begin{tabular}{c|ccc}
\hline
\textbf{Horizon} & \textbf{Centralized} & \textbf{Distributed (P-only)} & \textbf{Distributed (PI)} \\
\hline
1  & $1.82\times 10^{-9}$ & $1.90\times 10^{-1}$ & $1.02\times 10^{-2}$  \\
2  & $3.39\times 10^{-9}$ & $3.80\times 10^{-1}$ & $1.97\times 10^{-2}$  \\
3  & $5.23\times 10^{-9}$ & $4.51\times 10^{-1}$ & $2.82\times 10^{-2}$  \\
4  & $7.50\times 10^{-9}$ & $5.26\times 10^{-1}$ & $3.69\times 10^{-2}$  \\
5  & $9.94\times 10^{-9}$ & $5.80\times 10^{-1}$ & $5.08\times 10^{-2}$  \\
6  & $1.24\times 10^{-8}$ & $6.18\times 10^{-1}$ & $5.48\times 10^{-2}$  \\
7  & $1.53\times 10^{-8}$ & $7.11\times 10^{-1}$ & $5.66\times 10^{-2}$ \\
8  & $1.51\times 10^{-8}$ & $7.74\times 10^{-1}$ & $6.03\times 10^{-2}$  \\
9  & $1.62\times 10^{-8}$ & $8.21\times 10^{-1}$ & $6.45\times 10^{-2}$  \\
\hline
\end{tabular}
\end{table}

We benchmark centralized and distributed implementations in MATLAB (Intel i7, 16GB RAM) for the studied $20 \times 20$ grid. Centralized learning, based on $K = Y X^\dagger$ (with $X^\dagger$ denoting the Moore–Penrose pseudoinverse), completes in $\sim$15.2 ms. Distributed learning takes $\sim$0.3 ms for local computation and $\sim$0.1 ms for communication per iteration, totaling $\sim$400 ms. Although centralized learning appears faster for small datasets, it requires full data aggregation and becomes more computationally expensive as the number of snapshots increases. Moreover due to energy, privacy, bandwidth, or field-of-view constraints, continuous high-resolution data collection from a single agent is often physically infeasible in real-world scenarios.

In our simulations, we employ direct vectorization of the $20 \times 20$ grid as observables. This choice is natural for representing spatial phenomena such as pedestrian density and preserves locality in the lifted space. It enables interpretable modeling while maintaining computational tractability. Given our focus on distributed Koopman operator learning under distributed data access, this observable structure offers a practical and consistent approach for operator estimation across agents. We note that all agents observe temporally partitioned data from the same environment using a shared lifting function, rather than agent-specific observables.
\textcolor{black}{To further examine the role of the lifting dimension, we conducted an additional experiment in which the observable set was enriched by including second-order terms, effectively increasing the expressiveness of the lifted space.
Compared to the baseline lifting that uses only first-order observables, the enriched lifting yields a noticeably faster reduction of the distributed objective value.
In particular, the quantity $\frac{1}{p}\sum_{i=1}^p \|Y - K_i(t)X\|_F$ reaches the threshold $10^{-2}$ after approximately $600$ iterations for first-order lifting (Figure~\ref{fig:koopman_fit}), whereas the same level of accuracy is achieved after roughly $300$ iterations when second-order terms are included.
This behavior highlights a fundamental trade-off in Koopman-based modeling.
Richer lifting functions improve the linear representability of nonlinear dynamics, leading to faster convergence in terms of iteration count.
However, this improvement comes at the cost of increased lifted dimension, which directly impacts memory requirements, local computation, and inter-agent communication.
In the distributed setting, where each agent must exchange lifted quantities, the use of higher-order observables approximately doubles both the local computation time and the communication load per iteration.
From a centralized perspective, increasing the lifting dimension also amplifies the computational burden of forming and inverting large data matrices, potentially offsetting the gains in convergence speed.
These results indicate that the choice of lifting structure should balance model expressiveness against computational and communication constraints, particularly in distributed and resource-limited scenarios.
The proposed framework accommodates such trade-offs naturally, allowing practitioners to tailor the lifting complexity to the available resources and desired convergence behavior.}
\textcolor{black}{\begin{remak}\label{remark:scalability}
In Algorithm~\ref{algo}, each agent $i$ maintains a local Koopman matrix $K_i\in\R^{n\times n}$ (with $n$ the lifted dimension) and exchanges it with its neighbors at each communication round; consequently, the per-iteration communication payload scales with the number of transmitted real values, i.e., on the order of $n^2$ per neighbor. 
In the simulation setting considered here, the lifting is a direct embedding of a $20\times 20$ intensity map, hence $n=400$ and $K_i$ contains $1.6\times 10^5$ entries, which remains manageable for episodic learning with sparse neighborhood communication.
The intended regime of the proposed method is sensing-constrained scenarios where the lifted dimension is moderate and sequential observations motivate distributed computation without centralized data aggregation. 
At the same time, we note that as the lifted dimension $n$ increases, exchanging full dense matrices imposes higher communication requirements, which naturally motivates consideration of additional structure in the Koopman representation for each agent.
\end{remak}}
\begin{remak}\label{remark:tuning}
The convergence of Algorithm~\ref{algo} is driven by the spectrum of $\bM$ in~\eqref{def:bM}, shaped by $k_P$, $k_I$:
\begin{itemize}
    \item \textbf{$k_P$:} enhances consensus by penalizing disagreement. Too high may suppress local fitting.
    \item \textbf{$k_I$:} reinforces long-term agreement but risks instability unless $\alpha$ is small.
    \item \textbf{$\alpha$:} must satisfy $\alpha < \alpha_{\max}$ from~\eqref{def:alpha_max}. Larger $k_P$, $k_I$ reduce the spectral radius of $\bM$, increasing $\alpha_{\max}$ and improving the convergence rate governed by $\rho_{\max}$ in~\eqref{def:rho_max}. Set $\alpha = \theta \alpha_{\max}$ with $\theta \in (0.3, 0.6)$.
    \item \textbf{Practice:} use moderate gains, compute $\alpha_{\max}$, and set $\alpha = 0.5 \alpha_{\max}$. To accelerate, increase $k_P$; for stability, decrease $\alpha$ or $k_I$. Choose $t_{\max}$ based on $\rho_{\max}$.
\end{itemize}
\end{remak}
\begin{remak}\label{remark:online}
In practice, each agent collects local data over a brief sequence of time steps, after which Algorithm~\ref{algo} is executed. This interpretation of sequential observations aligns with physical constraints such as frame-rate or communication delays in agent sensing. Figure~\ref{fig:koopman_fit} shows that the distributed Koopman operators $K_i$ achieve low reconstruction error after around 300 iterations, which occur on millisecond time scales that is significantly faster than the few-second intervals at which video snapshots (e.g., of pedestrian motion) typically arrive. This separation ensures convergence is feasible before new data becomes available.
As confirmed in Figure~\ref{fig:koopman_error_heatmap}, the average operator $K_{\mathrm{ave}}$ provides accurate multi-step predictions even under decentralization, enabling online forecasting while new data is being acquired.
For example, as three agents collect the next $N = 9$ snapshots (each with $m_i = 3$), the current $K_i$'s can be used for short-term prediction. Once the data is available, the algorithm is rerun to refine the models, making the framework suitable for real-time deployment in applications such as crowd monitoring.
\end{remak}
\begin{remak}\label{remark:convergence}
Algorithm~\ref{algo} converges exponentially with rate governed by $\rho_{\max}$ in~\eqref{def:rho_max}. As shown in Figure~\ref{fig:koopman_fit}, approximately an accurate reconstruction is achieved after $\sim$300 iterations.
In online settings, where time and resources are limited, sub-optimal iterates can still be used for short-term prediction. Figure~\ref{fig:koopman_error_heatmap} shows that even intermediate solutions yield reliable forecasts.
As new data arrives, the algorithm can resume from the latest iterate, enabling continual refinement. This supports real-time deployment, where learning proceeds alongside sensing and control.
\end{remak}

\begin{figure}[!t]
    \centering
\includegraphics[trim=0.0cm 0.0cm 0.0cm 0.0cm, clip, width=0.9\textwidth]{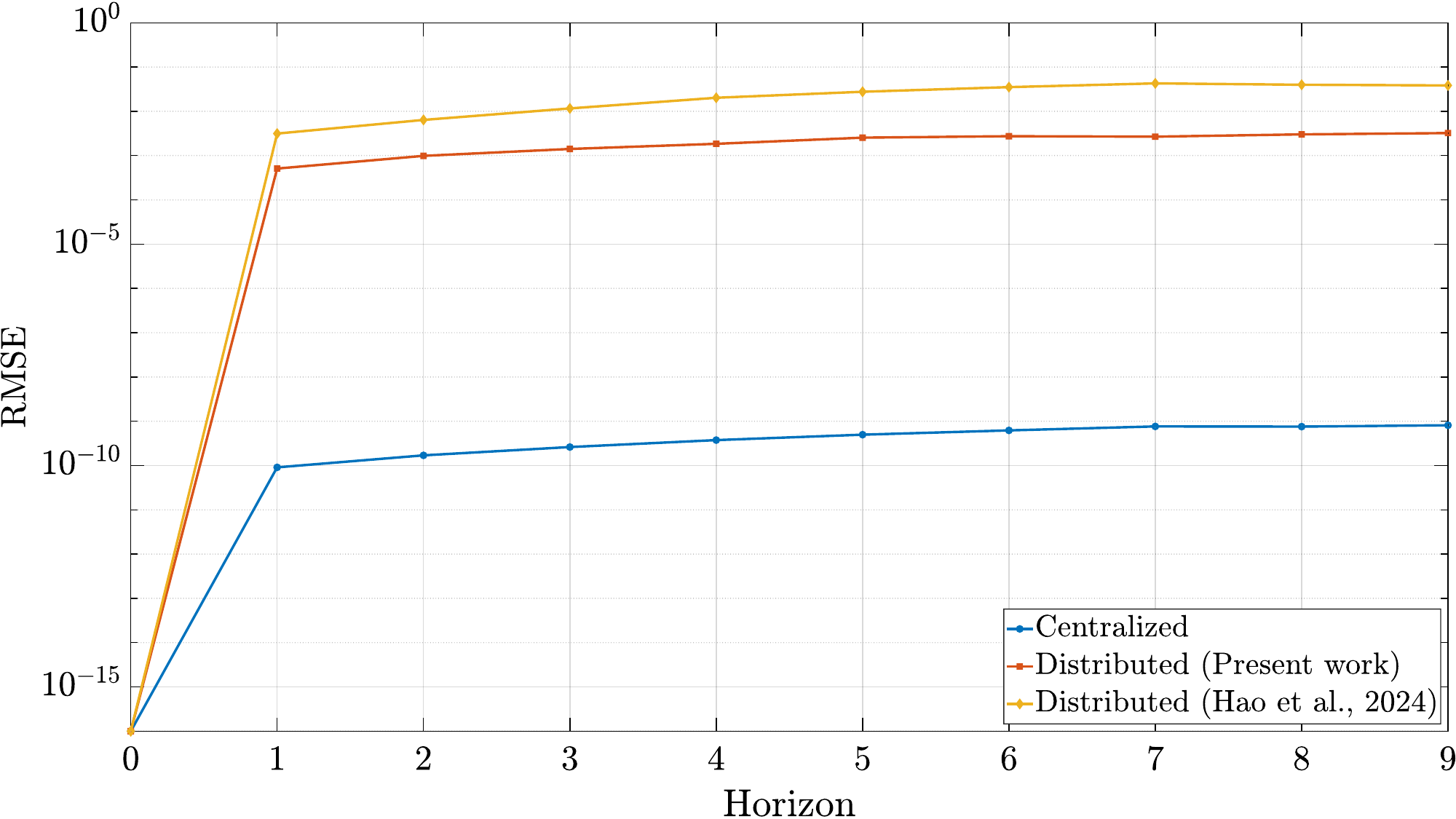}
    \captionsetup{font=footnotesize}
    \caption{
        {\textcolor{black}{RMSE over the prediction horizon for centralized and distributed Koopman learning methods: the proposed distributed method vs. the distributed baseline of~\citep{hao2024distributed222}
}}
    }
    \label{fig:prediction_rmse_comparison}
\end{figure}

\textcolor{black}{To contextualize the proposed method, we compare our distributed Koopman learning framework against the distributed approach introduced in~\citep{hao2024distributed222}, which represents one of the most closely related existing results.
Both approaches address the problem of learning Koopman operators from partial and distributed observations, making this comparison particularly relevant and fair.
To ensure consistency across methods, the lifting functions are kept fixed and identical for the simulation runs, and both distributed algorithms are executed for the same number of iterations over the same communication graph.
Moreover, since the present study focuses on autonomous prediction rather than control, the input-related components in~\citep{hao2024distributed222} are omitted, resulting in a formulation that aligns with our problem setup.}

\textcolor{black}{Prediction performance is evaluated using the root mean square error (RMSE) computed over the reserved set of unseen snapshots.
Specifically, for a prediction horizon, $h = 1,\ldots,9$, the RMSE is defined as
\[
\mathrm{RMSE}(h) = \frac{\|\hat{x}_h - x_h\|_2}{\sqrt{q}},
\]
where $\hat{x}_h \in \mathbb{R}^q$ denotes the predicted intensity field, $x_h$ is the corresponding ground-truth snapshot, and $q$ is the spatial dimension of the grid.
This metric quantifies the average per-cell prediction error and allows for a direct comparison of multi-step forecasting accuracy across methods.}

\textcolor{black}{The resulting RMSE curves over the prediction horizon are shown in Figure~\ref{fig:prediction_rmse_comparison}.
As expected, the centralized Koopman model achieves the lowest error across all horizons.
Both distributed methods exhibit stable multi-step prediction behavior on unseen data, indicating that the learned operators generalize beyond the training snapshots.
Among the distributed approaches, our proposed method consistently attains lower prediction error than the baseline distributed method of~\citep{hao2024distributed222} over the evaluated horizons, while preserving the distributed data acquisition and communication constraints.}

\section{Conclusion}
\label{Sec. 5}
This work proposed a distributed Koopman learning algorithm for modeling unknown nonlinear dynamics from sequential agent-level observations. Local Koopman models are constructed from lifted temporal data and refined via consensus to achieve consistent overall representations under distributed constraints.
\textcolor{black}{Simulation results demonstrate convergence and approximate model reconstruction using the proposed distributed learning method. In future, the learned Koopman operator can support real-time prediction of unknown motions, enabling proactive coordination and decision-making for dynamic obstacle avoidance in multi-agent scenarios.}

\appendix
\section{}
\label{app1}

The proof of Theorem~\ref{thm:col} relies on the following Lemma.    

	\begin{lemma}\label{lem:semi-Hurwitz}
		{\color{black}Consider a matrix
		\begin{equation}\label{def:tilde_M}
			\tilde\bM=\begin{bmatrix}
				-\bm X \bm X^\top-k_P\bL&\sqrt{k_I}\bL^{\frac{1}{2}}\\-\sqrt{k_I}\bL^{\frac{1}{2}}&0
			\end{bmatrix}.
		\end{equation}}
		It holds that $\Lambda(\bM)=\Lambda(\tilde\bM)$, where $\bM$ is  defined in \eqref{def:bM}.
	\end{lemma}
	
	\begin{proof}
		Since the matrix $\bL^{\frac{1}{2}}$ and $I$ commute, we use Schur complement and determinantal formula \citep[Chapter 0.8.5]{RAH-CRJ:12} to conclude
		\begin{align*}
			\det(sI-\bM)&=\det\left(\begin{bmatrix}
				sI+\bm X \bm X^\top+k_P\bL&-\bL\\k_II&sI
			\end{bmatrix}\right)\\
			&=\det(s(sI+\bm X \bm X^\top+k_P\bL)+k_I\bL)\\
			&=\det\left(\begin{bmatrix}
				sI+\bm X \bm X^\top+k_P\bL&-\sqrt{k_I}\bL^{\frac{1}{2}}\\\sqrt{k_I}\bL^{\frac{1}{2}}&sI
			\end{bmatrix}\right)\\
			&=\det(sI-\tilde\bM).\hspace{4cm}\qedhere
		\end{align*}
	\end{proof}

\begin{proof}[Proof of Theorem~\ref{thm:col}]
   Similar to the $\bK$ defined in \eqref{def:bK}, define
		\begin{equation*}
			\bR:=\begin{bmatrix}
				R_1&R_2&\cdots&R_p
			\end{bmatrix}\in\R^{n\times np}.
		\end{equation*}
	The update law \eqref{algo1_2} and the initial state $R_i(0)=0_{n\times n}$ imply the existence of $\tilde{\bR}(k)\in\R^{n\times np}$ such that $\bR(k)=\frac{1}{\sqrt{k_I}}\tilde{\bR}(k)\bL^{\frac{1}{2}}$ for all $k\in\N$. Therefore, the update law \eqref{algo1} can be rewritten as
	\begin{equation}\label{algo1_compact}
		\begin{bmatrix}
			\bK^+&
			\tilde{\bR}^+
		\end{bmatrix}=\begin{bmatrix}
		\bK&
		\tilde{\bR}
		\end{bmatrix}(I+\alpha\tilde{\bM}) +\alpha \begin{bmatrix}
		Y \bm X^\top&0
		\end{bmatrix},
	\end{equation}
	where we recall that the matrices $\bm X,\tilde{\bM}$ are defined in \eqref{def:bB}, \eqref{def:tilde_M}. 
    Meanwhile, it follows from the KKT condition \citep{Boyd_Vandenberghe_2004} that for any optimal solution $\bK^*\in\R^{n\times np}$ of the constrained problem \eqref{op_1}, there exists $\Lambda^*\in\R^{n\times np}$ such that
	\begin{subequations}
		\begin{align*}
			-(Y-\bK^*\bm X)\bm X^\top + \Lambda^*\bL&=0,\\
			\bK^*\bL&=0.
		\end{align*}
	\end{subequations}

\noindent
	These equations can be compactly written as
	{\color{black}\begin{equation}\label{KKT_compact}
		\begin{bmatrix}
			\bK^*&-\frac{1}{\sqrt{k_I}}\Lambda^*\bL^{\frac{1}{2}}
		\end{bmatrix}\tilde\bM+\begin{bmatrix}
			Y \bm X^\top&0
		\end{bmatrix}=0.
	\end{equation}}

\noindent
	For each $k\in\N$, define $\bE(k):=\begin{bmatrix}
		\bE_K(k)&\bE_R(k)
	\end{bmatrix}$, where
	\begin{equation*}
		\bE_K(k):=\bK(k)-\bK^*, \quad{\color{black}\bE_R(k):=\tilde{\bR}(k)+\frac{1}{\sqrt{k_I}}\Lambda^*\bL^{\frac{1}{2}}.}
	\end{equation*}
	{\color{black}It follows from \eqref{algo1_compact} and \eqref{KKT_compact} that
	\begin{align}
		\bE^+&=\begin{bmatrix}
			\bK^+&\tilde{\bR}^+
		\end{bmatrix}-\begin{bmatrix}
		\bK^*&-\frac{1}{\sqrt{k_I}}\Lambda^*\bL^{\frac{1}{2}}
		\end{bmatrix}\nonumber\\
		&=\begin{bmatrix}
			\bK&
			\tilde{\bR}
		\end{bmatrix}(I+\alpha\tilde{\bM})+\alpha \begin{bmatrix}
		Y \bm X^\top&0
		\end{bmatrix}
		\quad-\begin{bmatrix}
		\bK^*&\frac{1}{\sqrt{k_I}}\Lambda^*\bL^{\frac{1}{2}}
		\end{bmatrix}\nonumber\\
		&=\begin{bmatrix}
			\bK&
			\tilde{\bR}
		\end{bmatrix}(I+\alpha\tilde{\bM})-\alpha\begin{bmatrix}
		\bK^*&-\frac{1}{\sqrt{k_I}}\Lambda^*\bL^{\frac{1}{2}}
	\end{bmatrix}\tilde\bM \quad-\begin{bmatrix}
			\bK^*&\frac{1}{\sqrt{k_I}}\Lambda^*\bL^{\frac{1}{2}}
		\end{bmatrix}\nonumber\\
		&=\begin{bmatrix}
			\bK&
			\tilde{\bR}
		\end{bmatrix}(I+\alpha\tilde{\bM})-\begin{bmatrix}
			\bK^*&-\frac{1}{\sqrt{k_I}}\Lambda^*\bL^{\frac{1}{2}}
		\end{bmatrix}(I+\alpha\tilde\bM)\nonumber\\
		&=\bE(I+\alpha\tilde{\bM}),\label{update_E}
	\end{align}}
which is a matrix-valued linear time-invariant system. Hence, we only need to analyze the stability properties of the matrix $I+\alpha\tilde{\bM}$.
	For any eigenvalue $\lambda'$ of $I+\alpha\tilde{\bM}$, $\lambda'=1+\alpha\re(\lambda)+\alpha\im(\lambda)j$, where $\lambda$ is an eigenvalue of $\tilde\bM$. Hence
	\begin{equation*}
		|\lambda'|^2=(1+\alpha\re(\lambda))^2+\alpha^2\im(\lambda)^2=1+2\alpha\re(\lambda)+\alpha^2|\lambda|^2.
	\end{equation*}
	Meanwhile, by \citep[Lemma 7]{SL:24-arXiv},	$\tilde\bM$ is semi-Hurwitz; that is, all its eigenvalues either have negative real parts, or are $0$ and non-defective (algebraic multiplicity equals to geometric multiplicity) Additionally, because of Lemma~\ref{lem:semi-Hurwitz} and the assumption $\alpha<\alpha_{\max}$, where $\alpha_{\max}$ is defined in \eqref{def:alpha_max}, we either have $\lambda=0$ and non-defective so that $\lambda'=1$ and non-defective, or $\re(\lambda)<0$ and $|\lambda'|<1$. This implies that $I+\alpha\tilde{\bM}$ has a Jordan decomposition of the form
	\begin{equation}\label{JD}
		I+\alpha\tilde{\bM}=\begin{bmatrix}
			V_1&V_0
		\end{bmatrix}\begin{bmatrix}
		J_1\\&I
		\end{bmatrix}\begin{bmatrix}
		U_1\\U_0
		\end{bmatrix}=:VJU,
	\end{equation}
	such that $UV=I$, $J_1\in\R^{n'\times n'}$ for some $n'<2np$ is a block-diagonal matrix of Jordan blocks. Moreover, $J_1$ is Schur (all eigenvalues are in the open unit disk) and the spectral radius of $J_1$ is $\rho_{\max}$, defined in \eqref{def:rho_max}.
	Since $\rho>\rho_{\max}$, $\frac{1}{\rho}J_1$ is still Schur and {\color{black}hence by the discrete-time Lyapunov equation, there exists a symmetric positive definite matrix $P\in\R^{n'\times n'}$ such that 
	\begin{equation}\label{Lyapunov}
		\frac{1}{\rho^2}J_1PJ_1^\top- P\prec 0,
	\end{equation}
	where $\prec 0$ means the matrix is negative definite.}
	For each $k\in\N$, denote $\bF_1(k):=\bE(k)V_1, \bF_0(k):=\bE(k)V_0$.

    \noindent
    It follows from \eqref{update_E} and \eqref{JD} that
	\begin{align*}
	\bF_0^+&=\bE^+V_0 
	=\bE\begin{bmatrix}
		V_1&V_0
	\end{bmatrix}\begin{bmatrix}
		J_1\\&I
	\end{bmatrix}\begin{bmatrix}
		U_1\\U_0
	\end{bmatrix}V_0\\
	&=\bE V_0=\bF_0.
	\end{align*}
	Hence $\bF(k)=\bF(0)$. Similarly, 
	\begin{align*}
		\bF_1^+&=\bE^+V_1
		=\bE\begin{bmatrix}
			V_1&V_0
		\end{bmatrix}\begin{bmatrix}
			J_1\\&I
		\end{bmatrix}\begin{bmatrix}
			U_1\\U_0
		\end{bmatrix}V_1\\
		&=\bE V_1J_1=\bF_1J_1.
	\end{align*}
	Then, it further follows from \eqref{Lyapunov} that	
	\begin{align*}
	\Vert \bF_1^+P^{\frac{1}{2} }\Vert_F^2&=\trace( \bF_1^+P (\bF_1^+)^\top)
	=\trace( \bF_1J_1P J_1^\top\bF_1^\top)\\
	&\leq\rho^2\trace(\bF_1P\bF_1^\top)
	=\rho^2\Vert \bF_1P^{\frac{1}{2}}\Vert_F^2,
	\end{align*}
	where we have used the fact that the trace of positive semi-definite matrix is non-negative for the inequality above. 
	Hence, $\Vert \bF_1(k)P^{\frac{1}{2} }\Vert_F\leq \rho^k\Vert \bF_1(0)P^{\frac{1}{2}}\Vert_F$. In other words, $\bF_1(k)$ converges to $0$ exponentially with rate $\rho$. Because $\bE=\begin{bmatrix}
		\bF_1&\bF_0
	\end{bmatrix}U=\bF_1U_1+\bF_0U_0=\bF_1U_1+\bE V_0U_0$, we conclude that $\bE(k)$ converges to $\bE_\infty:=\bE(0)V_0U_0$ exponentially with rate $\rho$. In addition,
	\begin{multline*}
		\bE_\infty(I+\alpha\tilde{\bM})=\bE(0)V_0U_0\begin{bmatrix}
			V_1&V_0
		\end{bmatrix}\begin{bmatrix}
			J_1\\&I
		\end{bmatrix}\begin{bmatrix}
			U_1\\U_0
		\end{bmatrix}\\
		=\bE(0)\begin{bmatrix}
			0&V_0
		\end{bmatrix}\begin{bmatrix}
		J_1\\&I
		\end{bmatrix}\begin{bmatrix}
		U_1\\U_0
		\end{bmatrix}=\bE(0)V_0U_0=\bE_\infty,
	\end{multline*}
	which implies
	\begin{equation}\label{kernel}
		\bE_\infty\tilde{\bM}=0.
	\end{equation}
	Recall the definition of $\bE$. The solution of \eqref{algo1}, $\begin{bmatrix}
		\bK(k)&\tilde{\bR}(k)
	\end{bmatrix}$, converges to 
    $$\begin{bmatrix}
	\bK_\infty&\bR_\infty
	\end{bmatrix}:=\bE_\infty+\begin{bmatrix}
	\bK^*&-\frac{1}{\sqrt{k_I}}\Lambda^*\bL^{\frac{1}{2}}
	\end{bmatrix}.$$
	It then follows from \eqref{KKT_compact} and \eqref{kernel} that
	\begin{equation*}
		\begin{bmatrix}
			\bK_\infty&\bR_\infty
		\end{bmatrix}\tilde{\bM}+\begin{bmatrix}
		Y\bm X^\top&0
		\end{bmatrix}=0.
	\end{equation*}
	In other words, $\bK_\infty, \bR_\infty$ also satisfy the KKT condition of optimality for \eqref{op_1}. Hence, $\bK_\infty$ is an optimal solution for the problem \eqref{op_1}, and $\bK_\infty=\begin{bmatrix}
	K_\infty&K_\infty&\cdots&K_\infty
	\end{bmatrix}$, where $K_\infty\in\R^{n\times n}$ is an optimal solution for the problem \eqref{eq:frobenius_minimization}. This completes the proof.
\end{proof}
    %





 \bibliographystyle{elsarticle-harv} 
 \bibliography{cas-refs.bib}




\end{document}